\documentclass[twocolumn]{aastex62}

\usepackage{amsmath,amsfonts}

\usepackage{xcolor}

\shorttitle{Robust Registration of Astronomy Catalogs}
\shortauthors{Tian et al.}

\begin{document}

\title{Robust Registration of Astronomy Catalogs with Applications to \\ the Hubble Space Telescope}

\author{Fan Tian}
\affiliation{Dept.~of Applied Mathematics and Statistics, The Johns Hopkins University}

\author{Tam\'{a}s Budav\'{a}ri}
\affiliation{Dept.~of Applied Mathematics and Statistics, The Johns Hopkins University}
\affiliation{Dept.~of Computer Science, The Johns Hopkins University}
\affiliation{Dept.~of Physics and Astronomy, The Johns Hopkins University}

\author{Amitabh Basu}
\affiliation{Dept.~of Applied Mathematics and Statistics, The Johns Hopkins University}
\affiliation{Dept.~of Computer Science, The Johns Hopkins University}

\author{Stephen H. Lubow}
\affiliation{Space Telescope Science Institute}

\author{Richard L. White}
\affiliation{Space Telescope Science Institute}

\begin{abstract}

Astrometric calibration of images with a small field of view is often inferior to the internal accuracy of the source detections due to the small number of accessible guide stars. One important experiment with such challenges is the Hubble Space Telescope (HST). A possible solution is to cross-calibrate overlapping fields instead of just relying on standard stars. Following the approach of \citet{2012ApJ...761..188B}, we use infinitesimal 3D rotations for fine-tuning the calibration but devise a better objective that is robust to a large number of false candidates in the initial set of associations. Using Bayesian statistics, we accommodate bad data by explicitly modeling the quality, which yields a formalism essentially identical to an $M$-estimation in robust statistics. Our results on simulated and real catalogs show great potentials for improving the HST calibration, and those with similar challenges.

\end{abstract}

\keywords{catalogs --- astrometry --- methods: statistical}

\section{Introduction} 
\label{sec:intro}
With increasingly available observations from telescopes, astronomy has become one of the most data-intensive fields of study today. The introduction of high-resolution detectors in recent astronomical projects has led to a rapid growth in both data volume and data complexity. To fully utilize information from the vast datasets, it is then essential (and often has a great potential for new discoveries) to combine observations across multiple wavelengths, at varying time domains, and sometimes between different messengers. Over the last decade, studies in the field of catalog cross-matching have made significant progress using statistical and computational tools. \citet{2008ApJ...679..301B} introduced a reliable framework for symmetric cross-identification of multiple observations based on Bayesian hypothesis testing, which has provided superior results on handling astrometric uncertainties in simulations  \citep{2009ApJ...705..739H}. Their methods have also been successfully applied in several studies for cross-matching with unknown proper motions \citep{2010ApJ...719...59K}, to incorporate photometry of galaxies \citep{2014A&A...563A..14M}, or to study radio morphology \citep{2013ASPC..475...33F} and galaxy clustering \citep{2017A&C....20...83M}. A review of methods is also available in \citet{2015AnRSA...2..113B}. More recent studies have also introduced combinatorial optimization methods for cross-identifying associations for 2-way matching \citep{2016AJ....152...86B} and for N-way matching \citep{2019ApJ...870...51S}. While the above studies have opened a door for developing new systems and algorithms to address many different problems, new challenges are presented each day to many astronomers for various scientific demands. Among others, a particularly challenging practice arises in cross-matching small images such as those taken by the Hubble Space Telescope (HST).  

Unlike large survey projects such as the Sloan Digital Sky Survey \citep[SDSS;][]{2000AJ....120.1579Y} designed to provide a catalog, the HST is not used as a survey telescope in general. For more than twenty-five years, the HST has been operated under many independent programs targeting specific astronomical objects or sky regions using different detectors. The resultant HST data is a diverse collection of information from all observations made in the past including overlapping exposures at different angles and observations detected in different filters at different timelines. Cross-matching Hubble images to register the detected sources to a known catalog is more than matching nearby sources as studied in the aforementioned research. It also involves a step of positional adjustment of the images to better align the overlapping sources before matching.

While traditional image registration using the World Coordinate System \citep[WCS;][]{2002A&A...395.1061G} standard transformations and a guide star catalog is adequate for typical exposures, and automatic tools are available for blind registration to surveys \citep{2010AJ....139.1782L}, deep and narrow exposures, such as those taken by the HST, continue to be challenging due to the limited number of calibrators in the field of view. 
\citet{2012ApJ...761..188B} introduced a novel efficient method driven by the challenges of building the Hubble Source Catalog \citep[HSC;][]{2016AJ....151..134W}.
Using 3D rotations, they were able to cross-calibrate sources across the HST visits (telescope pointings) to obtain an improved relative astrometry. With the number of standard stars increased in the aligned images, there is an increased chance of further matching these astrometrically corrected images to the lower density, large reference catalogs.

To align the overlapping HST images, \citet{2012ApJ...761..188B} introduced a 3D infinitesimal rotation vector, which represents the axis and the angle of the rotation for an image. In the context of small corrections, the 3D rotation is also preferred over the traditional transformation performed on the tangent plane, since it avoids many expensive evaluations of the trigonometric functions. The shifts of the images are then determined by minimizing the separations between paired sources and calibrators that are close on the celestial sphere. This approach essentially arrives at the optimization of a quadratic cost function. The algorithm works effectively when the initial image offset is small, but the issue raises for large residuals that can overpower small values in estimation. The current solution to this problem in HSC is to pre-determine approximately matched pairs using the \textit{pre-offsets} method and a Bayesian likelihood comparison approach \citep{2016AJ....151..134W, 2012ApJ...761..188B}. 
A drawback of this method is that it uses a grid of limited resolution for obtaining the registration. In this study, we propose a new approach that is free from the step of pre-defining the set of nearly matched pairs and is free of grid resolution issues. To solve for the best transformation, we formulate a robust objective function that can tolerate a large number of erroneous associations in the initial set of candidate matches.

The paper is organized as follows. 
Section \ref{sec_method} presents the robust Bayesian approach and its connection to $M$-estimation with an illustration of some of the implementation considerations. 
Section \ref{sec_dis} discusses the results on the simulated and the real catalogs. 
Section \ref{sec_final} concludes the study.

\section{Methodology}
\label{sec_method}

The goal is to cross-register the source lists from multiple exposures, images or visits that have (even just partial) overlaps. We form groups of these source lists and work with their astrometric registration separately.
At the core of our iterative procedure is a simpler step that can correct the registration of a single source list (from one visit) onto a set of calibrators, whose direction is assumed to be perfectly known. This simpler problem is also very challenging in the presence of many false calibrators that are not actual guide stars. A robust treatment of that problem is the main focus of this paper. The iterative solution that cycles through the individual source lists using the candidate associations as calibrators has been discussed in detail by \citet{2012ApJ...761..188B}.
 
\subsection{Correcting to Calibrators}

Before carrying out the alignment of sources and calibrators, we need to first determine a set of initial associations. These candidate matches are obtained by considering for every source in the catalog, all calibrators within an angular separation of $R$; this would depend on the astrometry of the relevant catalogs. We index all these possible associations by $q \in \{1, \ldots, N\}$. 
Let $\boldsymbol{r}_q$ represent the direction (3D unit vector) of the source and its corresponding calibrator's direction $\boldsymbol{c}_q$ in the $q$th association. Note that the same source in a catalog will typically appear in multiple associations, i.e., different $q$, depending on $R$. 
An infinitesimal 3D rotation given by some $\boldsymbol{\omega}$ vector yields the updated direction
\begin{equation}
\label{eq:transformation}
\mbox{$\boldsymbol{r}_q^{\prime}(\boldsymbol{\omega}) = \boldsymbol{r}_q+\boldsymbol{\omega}\times \boldsymbol{r}_q$}\,.
\end{equation}

The second term is a small offset perpendicular to the original $\boldsymbol{r}_q$, which preserves the normality of $\boldsymbol{r}_q'$, with $\boldsymbol{r}_q'$ approximately an unit vector after transformation. The shift depends on the length and direction of $\boldsymbol{\omega}$. When $\boldsymbol{\omega}$ is co-linear with the image center, the transformation is a rotation around that axis, but if it is perpendicular to it, we get a simple shift in the tangent plane. The power of the above formula comes from the fact that it can describe both rotation and translation in the plane of the image, while being linear in the components of $\boldsymbol{\omega}$ vector. 

\color{black}

\subsection{Bayesian Formalism}

The approach introduced here considers the possibility that a (potentially large) fraction of the initial associations do not correspond to the same astronomical object.
Our goal is to simultaneously find the optimal transformation and identify the true associations.

Beyond the parameter of interest $\boldsymbol{\omega}$, we introduce a set of binary variables \mbox{$\beta\!=\!\{\beta_q\}$} to represent the two possible states for each candidate association $q$, which we will refer to as ``good'' and ``bad''. The former is an association belonging to the same object, but they are mixed with many false or bad ones. 
In addition, let $\gamma$ denote the probability of an association being ``good'', i.e., truly corresponding to the same object. 
First we formulate the problem in the Bayesian framework defining the likelihood function and the prior density function. Next we discuss the marginalization of the posterior to find a robust estimate of the calibration.
Let $p(\boldsymbol{\omega},\beta,\gamma)$ represent the joint prior probability density function (PDF) of all model parameters. 
Given a set of source-calibrator pairs data \mbox{$D\!=\!\{(\boldsymbol{r}_q, \boldsymbol{c}_q)\}$} and the likelihood function $p(D\lvert\boldsymbol{\omega},\beta,\gamma)$, one can obtain the posterior PDF on $\boldsymbol{\omega},\beta,\gamma$ using Bayes' rule. 
After marginalizing over the nuisance parameters $\beta$ and $\gamma$ we get the posterior distribution on $\boldsymbol{\omega}$ (conditioned on the data set):
\begin{equation}\label{eq:post1}
    p(\boldsymbol{\omega}|D)\propto \int\!d\gamma \sum_{\beta}\,p(\boldsymbol{\omega}, \beta, \gamma)\,p(D|\boldsymbol{\omega},\beta,\gamma)\,.
\end{equation}
We note that the sum over $\beta$ considers all possible combinations of the individual $\beta_q$ components. If there are $N$ candidate associations, there are $2^N$ possibilities to consider, which is typically a very large number even for images with only a few thousand detections. Hence, the direct evaluation of the above sum would be computationally prohibitive due to the combinatorially large number of terms, but in our case it can be done analytically.

On the one hand, the joint prior density function 
has simplified dependencies because the transformation can be assumed to be independent from the calibrators, so the product of the conditional probabilities become
\color{black}
\begin{equation}
    p(\boldsymbol{\omega},\beta,\gamma) = 
    p(\boldsymbol{\omega})\,p(\gamma)\,p(\beta|\gamma).
\end{equation}
On the other hand, the likelihood function $p(D|\boldsymbol{\omega},\beta,\gamma)$ depends only on $\boldsymbol{\omega}$ and $\beta$, which we can write as
\begin{equation}
    L(\boldsymbol{\omega},\beta) = \left[\prod_{q:\,\beta_{q}=1} \ell_{q}^G (\boldsymbol{\omega})\right]
    \left[\prod_{q:\,\beta_{q}=0} \ell_{q}^B (\boldsymbol{\omega})\right]
\end{equation}
where $\ell_{q}^{G}(\boldsymbol{\omega})$ and  $\ell_{q}^{B}(\boldsymbol{\omega})$ are the ``good'' and ``bad'' member likelihood functions respectively. 

A natural choice for the member likelihood function which describes the directional uncertainty is the Von Mises-Fisher distribution \citep{Fisher295} --- a spherical analogue to the Gaussian. Alternatively, one can also model the bivariate normal distribution on a unit sphere with a Kent distribution \citep{1993sasd.book.....F, ley2017}. 

In our approach, we choose to model with the Von Mises-Fisher distribution. For the observed direction $\boldsymbol{x}$ and the direction of the mode $\boldsymbol{r}$, the PDF of the Fisher distribution is defined as
\begin{equation}
    F(\boldsymbol{x};\boldsymbol{r},\kappa) = \frac{\kappa}{4\pi\sinh\kappa}\exp (\kappa\,\boldsymbol{r} \boldsymbol{x})
\end{equation}
with the compactness parameter $\kappa$. For the typical small astrometric uncertainty, one can use the flat-sky approximation locally where a (bivariate) normal distribution describes the uncertainty in the tangent plane, in which case \mbox{$\kappa\!=\!1/\sigma^2$}, where $\sigma$ is the astrometric uncertainty. For a known point-spread function (PSF) associated with each point source, we can take $\sigma$ as the individual positional error for each source, which is typically a fraction of the PSF width. Alternatively, we use a constant systematic positional error for all sources, ignoring any variation in astrometric accuracy in this case.

The ``good'' member likelihood function thus can be written as
\begin{equation}
\ell_{q}^{G}(\boldsymbol{\omega})=F\!\left(\boldsymbol{c}_{q};\boldsymbol{r}_{q}^{\prime}(\boldsymbol{\omega}),\kappa\right)
\end{equation}
where \mbox{$\boldsymbol{r}_{q}^{\prime}(\boldsymbol{\omega})$} describes the transformation using infinitesimal rotation, see Equation~\eqref{eq:transformation}.
The ``bad'' member likelihood function can be assumed to be isotropic, i.e., uniform over the field of the image,
\begin{equation}
\ell_{q}^{B}(\boldsymbol{\omega})=\frac{1}{\mathbf{A}}
\end{equation}
where $\mathbf{A}$ is the area of the image. 

Considering the prior probability on $\beta$ given $\gamma$ for the good and bad pairs explicitly such that
\begin{equation}
    p(\beta|\gamma) = \left[\prod_{q:\, \beta_{q} =1} \gamma \right]\left[\prod_{q:\, \beta_{q} =0} (1-\gamma) \right],
\end{equation}
the joint posterior probability distribution of $\boldsymbol{\omega}$ from Equation~\eqref{eq:post1} is then given as
\begin{align} \label{eq:post2}
\begin{split}
&    p(\boldsymbol{\omega}|D) \propto 
     p(\boldsymbol{\omega}) \times \\
&    \int\!\!d\gamma\, p(\gamma)\sum_{\beta}\left[\prod_{q:\, \beta_{q} =1} \gamma \ell_{q}^{G}( \boldsymbol{\omega}) \right]
    \!\!
    \left[\prod_{q:\,\beta_{q}=0}\!(1\!-\!\gamma)  \ell_{q}^{B}(\boldsymbol{\omega})\right].
\end{split}
\end{align}

We can analytically marginalize over all combinatorial possibilities described by recognizing that the sum over $\beta$ is a product of the mixture,
\footnote{Equation~\eqref{eq:post2} is also known as the Bernoulli Mixture Model which can be solved with maximum-likelihood type estimation using EM algorithm \citep{dempster1977}.}
\begin{equation} \label{eq:post3}
    p(\boldsymbol{\omega}|D)\propto 
    p(\boldsymbol{\omega})\!\!\int\!\!d\gamma\,p(\gamma)\!\prod_{q}\!\left[\gamma\ell_{q}^{G}(\boldsymbol{\omega})+
    (1\!-\!\gamma)\ell_{q}^{B}(\boldsymbol{\omega})\right]
\end{equation}
which is a much simpler formula to work with. Numerically, one can map out the marginal posterior using sampling methods such as Markov chain Monte Carlo (MCMC) or calculate the expectation value of $\boldsymbol{\omega}$. 

In practice, the product over $q$ in the above result is surprisingly insensitive to the (small) value of $\gamma$. 
In fact, a simple Dirac delta prior \mbox{$p(\gamma)\!=\!\delta(\gamma-\gamma_*)$} is suitable with reasonable choices of $\gamma_*$.
Since $\gamma_*$ is just the fraction of the true associations, we can estimate its value based on the input catalogs or even find its maximum likelihood estimate, see \citet{2015AnRSA...2..113B}. 

For example, with $N_1$ sources and $N_2$ calibrators, the crudest estimate is the minimum of the two divided by the total number of possible associations within some search radius $R$,
\begin{equation}\label{eq:gamma}
    \gamma_{*}=\frac{\min(N_1, N_2)}{N}\,.
\end{equation}
This is really an upper bound of $\gamma$ but is a suitable estimate for the following procedure.

Substituting the member likelihood functions and a fixed $\gamma_*$ yields an ``effective'' likelihood function

\begin{equation} \label{eq:maxfn}
    L^{\!*}\!(\boldsymbol{\omega})=\!
    \prod_{q}
    \left[\frac{\gamma_{*}}{2\pi\sigma^{2}}
    \exp\left\{\!-\frac{\left[\boldsymbol{c}_{q} - (\boldsymbol{r}_{q}\!+\!\boldsymbol{\omega}\!\times\!\boldsymbol{r}_{q})\right]^{2}}
    {2\sigma^{2}}\right\} +\frac{1\!-\!\gamma_{*}}{\mathbf{A}}\right]
\end{equation}
Where efficiency is imperative, such as the Hubble Source Catalog where hundreds of millions of detections appear somethings in thousands of overlapping visits drive the computation cost high, a simple shortcut is to maximize this function to obtain a solution for $\omega$ that is robust against the overwhelmingly large fraction of bad associations.

\begin{figure}
  \centering
  \includegraphics[width=0.9\columnwidth]{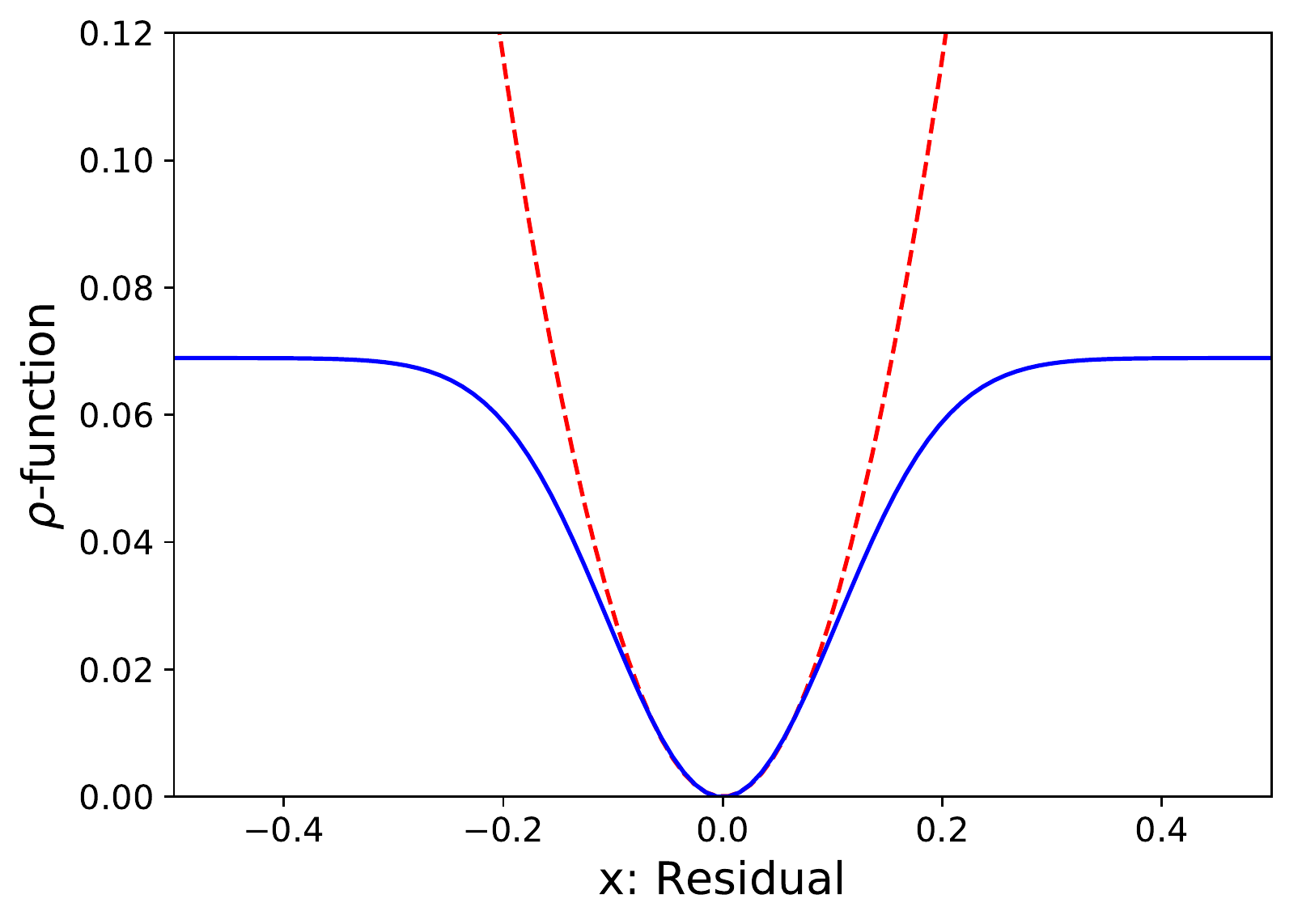}
  \caption{The robust $\rho$-function (\textit{solid blue line}) limits the influence of outliers in comparison to a quadratic objective (\textit{dashed red line}).}
  \label{fig:rho_fn}
\end{figure}

\subsection{Connection to \textit{M}-estimation}
When all pairs are ``good'', i.e. $\gamma_{*}=1$, Equation~\eqref{eq:maxfn} yields the least-squares problem as introduced in \citet{2012ApJ...761..188B}. As the fraction of good pairs decreases, the effective likelihood function gains heavier tails making the optimization more difficult. 
Estimation with the problem of high unbalancedness has been discussed in previous literature of \citet{brown2001}, with \citet{2006ApJ...652..610P} proposed a Bayesian approach for numerical computation. In this study, to find the optimum, we borrow ideas from robust statistics \citep{Huber1981, maronna2006} to reformalize our objective function in Equation~\eqref{eq:maxfn}.

Let the separation between the $q$-th source-calibrator pair to be $\boldsymbol{\Delta}_{q}\!=\!\boldsymbol{c}_{q}\!-\!\boldsymbol{r}_{q}$. For any given $\gamma_{*}$, instead of maximization, we minimize the negative logarithm of the likelihood function in Equation~\eqref{eq:maxfn} and thus arrive at the following objective function
\begin{equation} \label{eq:rho}
    \tilde{\boldsymbol{\omega}}= 
    \arg\min_{\boldsymbol{\omega}}\sum_{q}\,
    \rho\left(\frac{ \left|\boldsymbol{\Delta}_{q}-\boldsymbol{\omega}\times\boldsymbol{r}_{q} \right|}{\sigma}\right)
\end{equation}
\begin{equation*}
    \textrm{with} \quad
    \rho(x) = -\ln 
    \left(
    \frac{\gamma_{*}}{2\pi\sigma^2}\ e^{-x^2/2} \,+\, \frac{1\!-\!\gamma_{*}}{\mathbf{A}}
    \right).
\end{equation*}

As illustrated in Figure~\eqref{fig:rho_fn}, this $\rho$-function is quadratic for small residuals, but constant for large values - limiting the contribution of bad pairs to the objective. 
We note that $\rho$ is a function of $x^2$ only and this problem formally is much like $M$-estimation in robust statistics. 

As first introduced by \citet{huber1964}, an $M$-estimator is a maximum-likelihood type estimator, which minimizes a function of errors instead of the sum of their squares to achieve more robustness. For instance, in the problem of finding the best fit given a sequence of data points $(X_i, Y_i)$, a maximum-likelihood estimator $\hat{\theta}$ minimizes the sum of squared errors of
\begin{equation}
    \sum_{i} (Y_{i}- X_{i}\theta)^{2}
\end{equation}
for $Y_i=\theta X_i+\epsilon_i$, where  $\epsilon_i$ are normally distributed error terms and are assumed to be independent. With $M$-estimation, the estimator $\hat{\theta}$ is determined by minimizing a function of the errors
\begin{equation}
    \sum_{i} \rho(Y_{i}- X_{i}\theta)
\end{equation}
for $\rho$ to be a non-constant function. More detailed discussions on the choice and the properties of the $\rho$-function can be found in literature and textbooks of robust statistics \citep{huber1964, Huber1981, maronna2006}.

The solution exists requiring the gradient of the objective function equals to zero. Since no closed-form solution exists for an $M$-estimation, a general approach is to use an Iteratively Reweighted Least-Squares (IRLS) method \citep{maronna2006}. In this study, we solve the problem by iterating between 
(i) solving for $\tilde{\boldsymbol{\omega}}$ using $A\,\tilde{\boldsymbol{\omega}}=b$ 
with
\begin{equation}\label{eq:solve}
\begin{array}{ccc}
    A =\displaystyle \sum_{q} \frac{w_{q}}{\sigma^{2}}
    \left(I-\boldsymbol{r}_{q}\!\otimes\boldsymbol{r}_{q}\right) & \textrm{and} &
    b = \displaystyle \sum_{q}
    \frac{w_{q}}{\sigma^{2}}
    \left(\boldsymbol{r}_{q}\!\times\boldsymbol{c}_{q}\right) 
\end{array}
\end{equation}
assuming constant $w_q$ weights, 
and 
(ii) re-evaluating those weights based on the new $\tilde{\boldsymbol{\omega}}$ as
\begin{equation}\label{eq:wq}
    w_{q}=W\left(\frac{\left|\boldsymbol{\Delta}_{q}-\tilde{\boldsymbol{\omega}}\times\boldsymbol{r}_{q}\right|}{\sigma}\right)
\end{equation}
with
\mbox{$W(t)=\rho'(t)/t$}.

Equation~\eqref{eq:solve}%
\footnote{The operator $\otimes$ represents the dyadic (or outer) product of the 3D (column) vectors.} 
is very similar to the least-squares estimate \citep{2012ApJ...761..188B} but includes the new weighting scheme where the iterative update Equation~\eqref{eq:wq} enters.
We find this procedure converges quickly in practice.
\color{black}

\subsection{Practical Considerations}

In practice, large systematic offsets are expected to cause difficulties primarily for two reasons. 
When using big search radii, (1) the typical initial separation of sources will be significantly larger than the astrometric uncertainties,  and (2) the fraction of true associations will become tiny. 
First, if the angular separations are much larger than the astrometric uncertainty, which we denote by $\sigma$, the robust $\rho$-function will have small gradients and the weights can shrink to the values that are numerically indistinguishable from zero. This numerical issue could halt the iterative procedure. We alleviate this by using large $\sigma$ values at the beginning of the iterations if needed and let it converge to the true astronometry as the solutions improve.

Second, the large fraction of bad candidates can be tackled by a computationally efficient algorithm.
We recall that the 3D transformation described by $\boldsymbol{\omega}$ vector is able to simultaneously correct for translation and rotation in the tangent plane. For small field of views, which is the most relevant limit, even relatively large rotation will yield small displacements in the tangent plane, hence big corrections are primarily required for large shifts. This means that the initial difference vectors \mbox{$\boldsymbol{\Delta}_q$} 
will be typically clustered for the good associations. In fact, this is the reason why heuristic methods using 2D histograms were successful in the past \citep{2015ASPC..495...65L}. 
Instead of searching in circles of increasing radii (as we originally introduced the method with threshold $R$), we can look for the right candidates in larger and larger rings with fixed width. While the number of candidates still increase, the growth is not quadratic, only linear.

In this paper we adopt to use rings with widths of 10$\sigma$.
To eliminate the possibility of the true associations falling on the boundary of two rings, we also overlap the rings by half their widths considering only $\boldsymbol{\Delta}_{q}$ that fall in those rings.
Sequentially starting from the smallest ring (or even in parallel), we can proceed to find the best transformation in each ring, and test the quality of the transformation. If the solution converges on some random pairs, the number of these will be very small, because of the small probability of the noise producing a systematic pattern. For good associations, the initial $\gamma_*$ estimate will be closes to the mean of $\{w_q\}$ weights. 

\section{Discussion}
\label{sec_dis}
We focus on observations of the Hubble Space Telescope's Advanced Camera for Surveys in the Wide Field Channel \citep[HST/ACS/WFC;][]{ACS2018} and study simulated visits to test the limitations of the new methods before applying it to calibrating real data to the Gaia DR2 catalog \citep{2018A&A...616A...2L}.
The Hubble Legacy Archive (HLA) discussed in \cite{2006ASPC..351..406J, 2012ApJ...761..188B} has
sources lists that were produced by the DAOPhot \citep{1987PASP...99..191S} and the Source Extractor \citep{1996A&AS..117..393B} softwares on the combined (whitelight) images within each HST visit \citep{2008ASPC..394..481W}. 
Other than the source directions, the source lists of the HLA also provide information such as the orientations, magnitudes and morphology of the detected sources \citep{2008ASPC..394..478M}. 

\color{black}

\begin{figure*}
\centering
\includegraphics[width=0.8\textwidth]{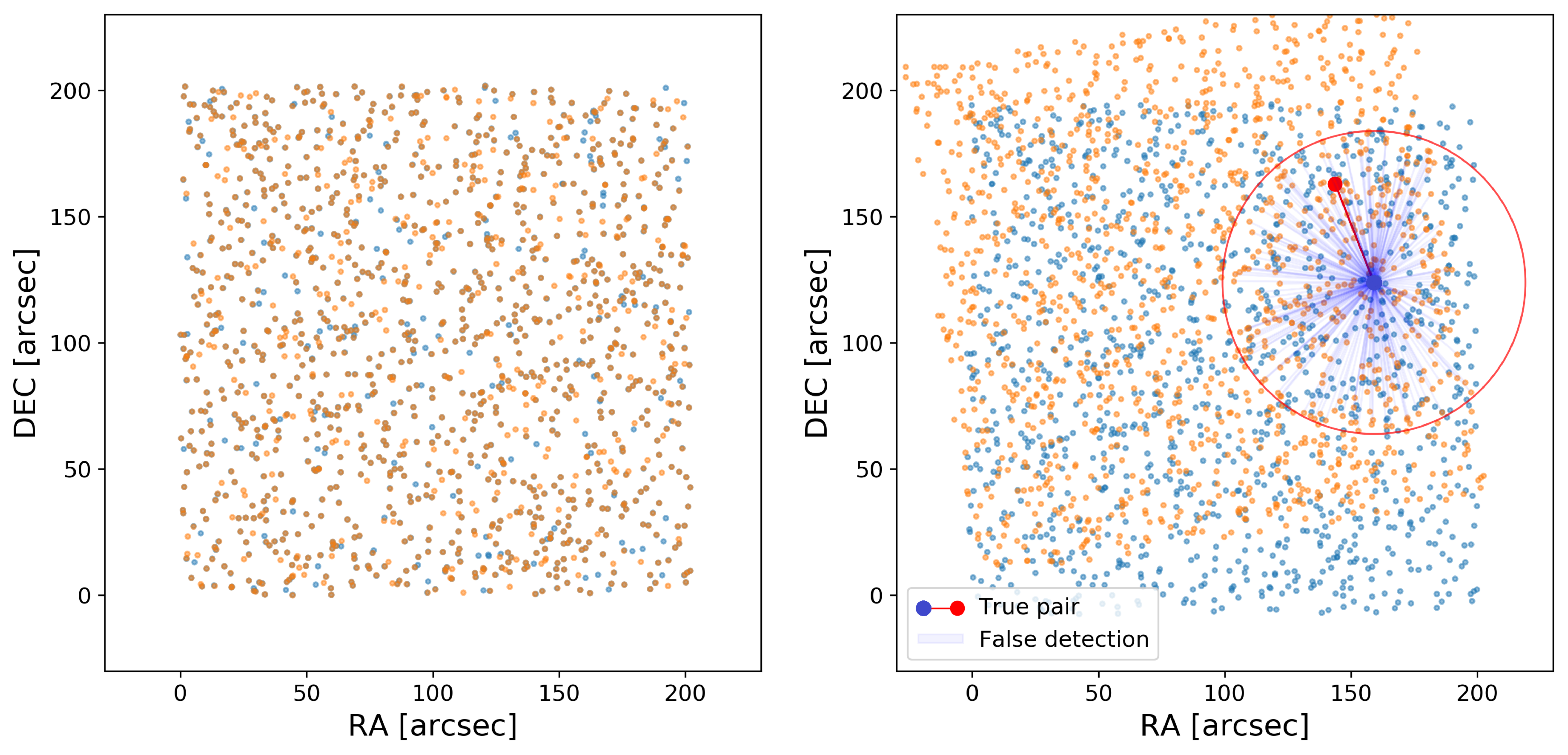}
\caption{Left: Two simulated catalogs (blue and orange dots) before applying a systematic transformation. Right: After transformation, the sources are not aligned. The large blue and red dots correspond to one randomly chosen object whose detections are connected with a red line. The circle encloses all false candidates matched to the large blue dot shown with blue thin lines.}
 \label{fig:2dcat}
\end{figure*}

\subsection{Simulated catalogs}

Our mock objects are point sources with random directions generated in a small field of view, which adapts the ACS/WFC image parameters of \mbox{$202''\!\times{}202''$} in size with approximately 1500 source detections. Each point is taken as an unit vector representing the pointing direction to the actual star coordinates on the celestial sphere. The astrometric uncertainty is taken with the HST positional accuracy of approximately 0.04 arcsec. Additional to the directional information, each object is assigned with a random stellar property $u_{01}$ drawn from an uniform distribution between 0 and 1.

From the mock universe, catalogs are generated in pairs by (1) assigning random perturbations to the mock objects with a chosen astrometric uncertainty $\sigma$; (2) selecting overlapping sources from the two catalogs by an interval constraint on the source property $u_{01}$; and (3) transforming the catalog pair with a random 3D rotation vector $\boldsymbol{\omega}$ drawn from a normal distribution. The estimation is then performed on the transformed catalog pairs, and our goal is to recover the $\boldsymbol{\omega}$ vectors applied. 

As an example to illustrate the simulation settings, the left panel of Figure~\eqref{fig:2dcat} is a 2D projection of a pair of the generated catalogs before transformation (point sources colored in blue and orange respectively). The right panel of Figure~\eqref{fig:2dcat} represents the same catalog pair after an $\boldsymbol{\omega}$ vector is applied. Additionally, Figure~\eqref{fig:2dcat} (right panel) also shows the challenges presented for cross-matching the two catalogs when the image offset is large and a larger search radius is used. For instance, for the singled-out source detection in one catalog (highlighted blue center), comparing to the many bad matchings represented by the blue lines within the search radius $R$ (red circle), obviously there is in fact only one true matching in the other catalog (red dot) where this pair of sources corresponds to the same underlying object.

\subsection{Testing the Limits}
Applying the same simulation settings but with different rotation vectors, we illustrate the cross-matching results of our new method on a set of catalog pairs for different search radii and different image offsets in this section. As for comparison, the traditional least-squares method is also tested under the same conditions. The method accuracy is reported by comparing the average initial offset of the true pairs between the two catalogs to the average offset of the pairs after correction.

\begin{figure}
    \centering
    \includegraphics[width=0.87\columnwidth]{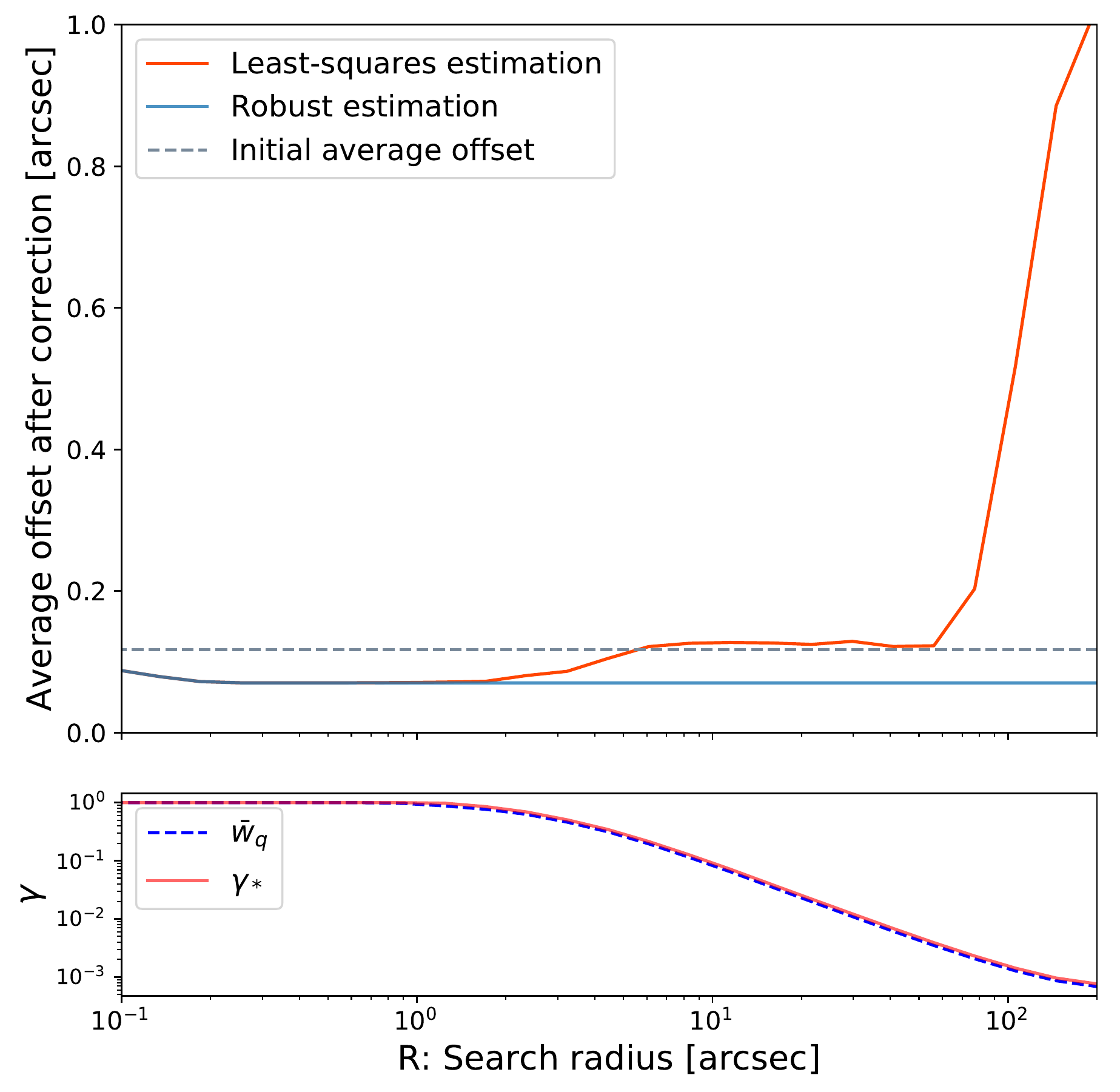}
    \caption{Comparison of least-squares estimation with the new robust estimation tested on two images with a small offset (\textit{grey dashed line}) and with increasing search radius. The top panel shows the offset of two images before and after correction. The bottom panel indicates the estimated $\gamma_{*}$ (red solid line) and computed average weights $\bar{w}_q$ (blue dashed line) probability of good pairs.}
    \label{fig:comparison1}
\end{figure}

We compare both the least-squares method and our robust method in two ways. We first test the estimation accuracy affected by the choice of the search radius for images with a small initial offset. As most of the Hubble image offsets are tenths of an arcsec, this test is performed on two images with an initial offset approximately 0.1 arcsec for generality. Since the offset is small, instead of searching over rings, we pair all nearby observations within a single search radius of $R$ for maximum $R$ to be 200 arcsec. The estimation results are shown in Figure~\eqref{fig:comparison1}. Referring to the top panel of the plot, as $R$ increases, both methods recover the correct rotation for \mbox{$R\!<\!1$} arcsec. For \mbox{$R\!>\!1$} arcsec, the least-squares method starts to break down. Our new robust estimate, on the other hand, can find the accurate rotation vector under large $R$. Here we also note that the slight initial decline in the estimation accuracy between 0.1 and 0.2 arcsec search radii is due to that we have not included a sufficient number of true pairs when using a smaller search radius than the image offset. 

The bottom plot of Figure~\eqref{fig:comparison1} represents the measurement of $\gamma$ of our robust method. The green line represents the average of the weights $w_q$ given in Equation~\eqref{eq:wq} and the red line is the estimated probability of the good pairs, $\gamma_*$, from Equation~\eqref{eq:gamma}. This plot shows that $\gamma_*$ is consistent with the fraction of the true associations determined in estimation when we recover the correct rotation. This also reinforces that our choice of the the good $\boldsymbol{\omega}$ estimates in rings with mean of $w_{q}$ approximately $\gamma_{*}$ is practical.

\begin{figure}
    \centering
    \includegraphics[width=0.87\columnwidth]{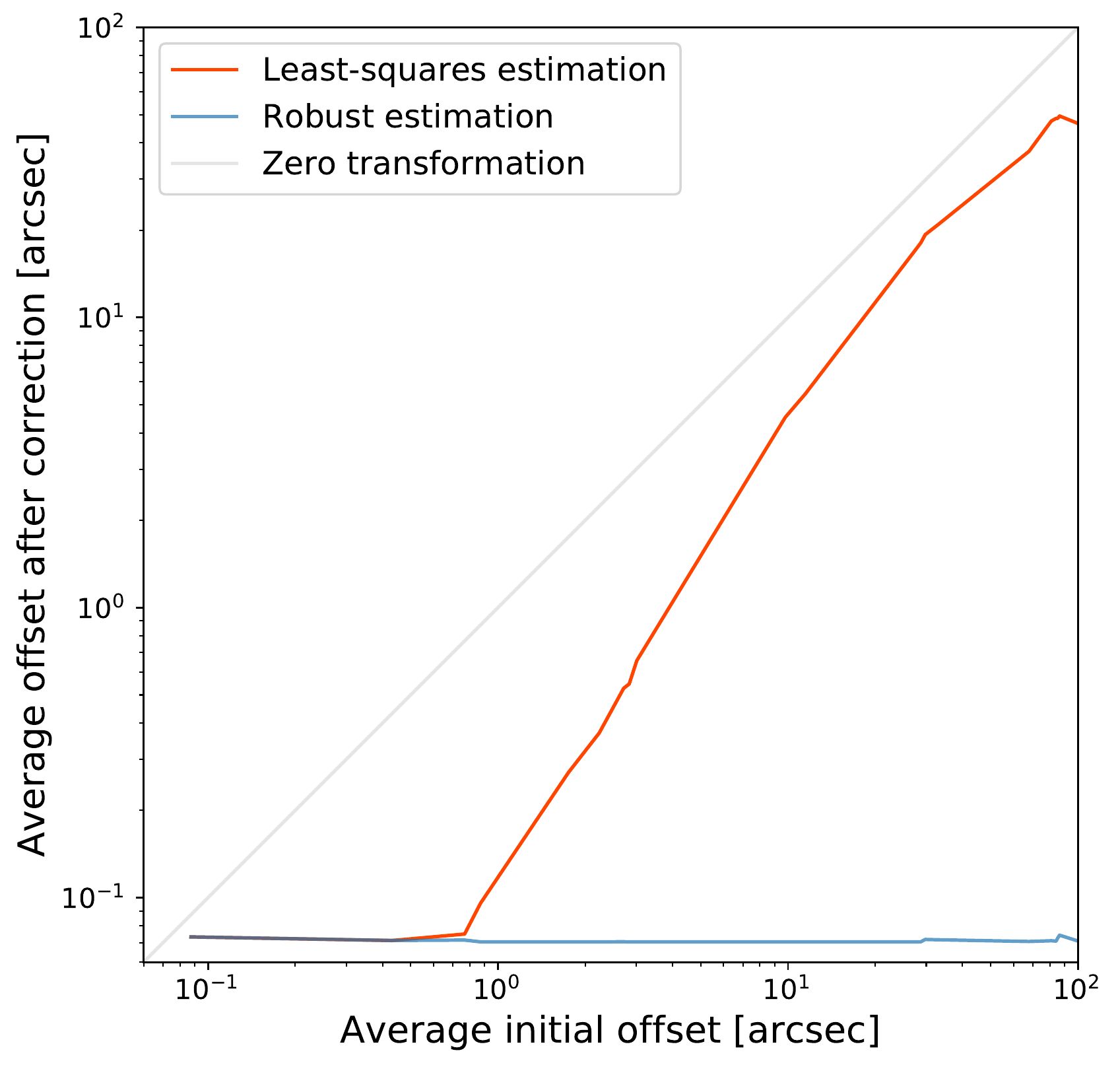}
    \caption{Comparison of the two methods with increasing average initial offset of images. The robust method is successful for images with large initial offset beyond 100 arcsec. The least-squares method corrects the astrometry for offsets less than 0.4 arcsec. Both estimations are performed on pairs within the same rings.}
    \label{fig:comparison2}
\end{figure}

Furthermore, we compare the two methods on images with increasing initial offsets and test for the method limitations. For efficient testing, here we applied the maximum upper bound of the search radii to be 10\% more than the maximum initial offset of the ground truth in each test. Since the least squares method is most accurate when the initial set of pairs are approximately matched. To draw a fair comparison of the two methods, we first determine the best rotation vectors in the optimal rings using our robust method, and obtain the set of pairs from the corresponding ring. This is also the set of most likely mathchings over all rings. The least squares estimation is then performed on the likely matched pairs in the optimal rings. The estimation results for both methods are reported as shown in Figure~\eqref{fig:comparison2}, and the grey solid line is the zero transformation line regarded as a reference. The correct rotation is only obtained when the average offset after correction is approximately $\sigma$. Any other estimate result deviates from $\sigma$ closes to the reference line is considered as an inaccurate solution. Refererring to the plot, we see that for images with a small initial offset ($<$ 0.4 arcsec), both our robust estimate and the least-squares estimate correct the astrometry to approximately $\sigma$. As the initial offset increases to above 0.4 arcsec, the least-squares algorithm fails to find a correction. This finding also coincides with the current limitation to the least-squares algorithm implemented in HSC. On the other hand, our robust estimate is accurate for large offsets beyond 100 arcsec. 

\begin{figure*}
\centering
\includegraphics[width=0.8\textwidth]{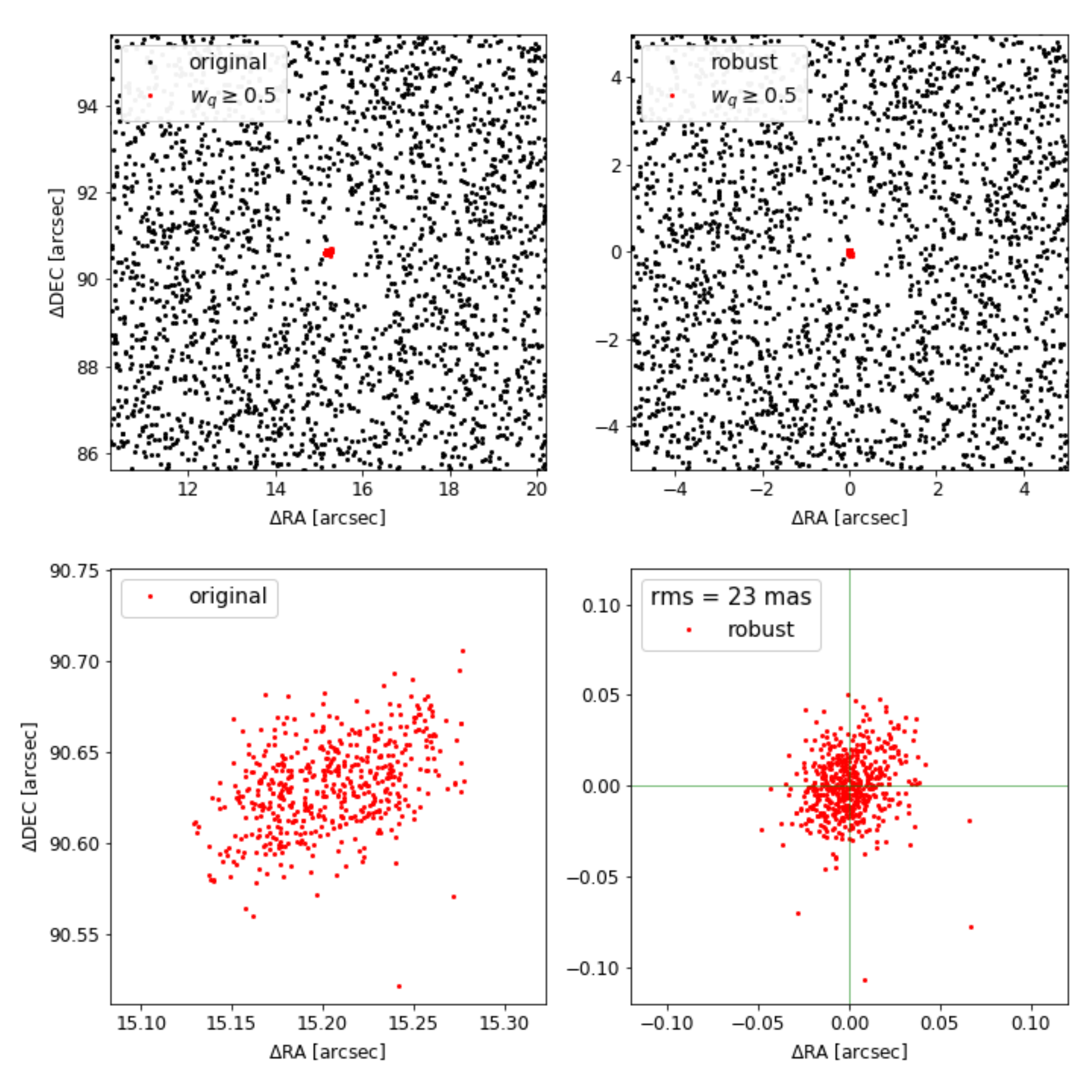}
\caption{Differences in the sky coordinates across the Hubble and the Gaia sources. Each point corresponds to the RA and Dec coordinate differences in a particularly challenging field. The left panels show the offset vectors across  associations (within a search radius) in the original Hubble and Gaia catalogs. The right panels illustrate the same after the robust transformation is performed. The center red points correspond to the associations that our procedure flags as good matches, which are also shown in detail in the bottom panels. Despite an offset of over 90 arcsec between the two catalogs, we have not only found the correct associations but the overall transformation also makes the ``good'' matches a tighter scatter, see comparison in the bottom two panels.}
\label{fig:hla-gaia}
\end{figure*}

\subsection{Matching Hubble and Gaia}
The new method was first applied to the HLA and Gaia observations to register the Hubble visits on Gaia's reference frame. 
While most cases required small tweaks to the coordinates systems, occasionally the Hubble images are completely off target, where previous methods broke down.
A collection of tricks were used to first create a crude registration using the 2D histogram method mentioned earlier and an $\boldsymbol{\omega}$-transformation as a follow-up. 
Even then, some cases required manual intervention to find the best registration. 
With the novel robust method all cases are handle by a single procedure that outperforms the previous approach. 
Here we discuss a most challenging scenario when the Hubble pointing is off by more than 90 arcsec.

The example we include here is taken at a field located near the plane of the Milky Way approximately 8 degrees from the galactic center. We cross-match the HLA source list of approximately 600 detections to Gaia's 10,000 sources in the relevant area. 
We find over one million candidate associations using an angular separation threshold of 120 arcsec. Among these we expect to find less than 600 true matches, which makes the contamination extremely large.

We approach the solution dividing the candidates into rings the same way we saw it in the simulations. The attempts to solve for a robust $\boldsymbol{\omega}$ fail in smaller rings (very low sum of weights) but a clear signal indicates a good solution at a ring just beyond 90 arcsec separations.
Figure~\eqref{fig:hla-gaia} illustrates the original offset vectors (left panels) and the residual differences after the robust transformation (right panels). 
Simultaneously, our procedure selects the ``good'' associations shown in red, which are highlighted in the bottom panels. 
Looking at these in detail we notice that not only the enormous systematic offset is removed but the scatter becomes significantly tighter, which indicates that we needed needed more than a simple translation to obtain a better astrometry.

\section{Summary}
\label{sec_final}
In this study, we have proposed a novel mathematical approach based on Bayesian and robust statistics for cross-registering astronomical catalogs tailored to observations with a small field of view. Our preliminary study on simulations to the HST observations has shown promising results on improving the astrometric accuracy over the state-of-art method. The new method successfully recovers transformations with a high astrometric accuracy under large search radii and even when the initial image offsets are very large. Unlike most previous methods, our new approach requires no pre-determination of the approximately matched pairs which also solves the grid resolution issues in the existing approaches. In addition to the simulation results, we have illustrated the power of our new method in a complex real scenario of cross-registering the Hubble images to the Gaia catalog. This also shows a potential of our robust method for improving the HSC astrometry by cross-matching to the Gaia DR2 in future. 

\section*{Acknowledgements}

The authors thank the anonymous reviewer for a thorough and insightful report. 
A.B. gratefully acknowledges support from NSF grant CMMI1452820 and ONR grant N000141812096.
T.B. gratefully acknowledges support from NSF Grant AST-1412566 and NASA via the awards NNG16PJ23C and STScI-49721 under NAS5-26555.

\end{document}